\documentclass[11pt, a4paper]{article}
\pdfoutput=1

\usepackage[height=8.85in,width=6.75in]{geometry}

\usepackage{chngcntr}
\counterwithin{equation}{section}

\usepackage{graphicx,rotating}     
\usepackage[bookmarksopen,colorlinks=true,linkcolor=light_blue,
citecolor=light_pink,urlcolor=dark_red,linktoc=all]{hyperref}

\usepackage{amsmath}
\usepackage{mathtools}
\usepackage{amsfonts}
\usepackage{amssymb}
\usepackage{slashed}
\usepackage{braket}
\usepackage{bm}
\usepackage{bbm}
\usepackage{cite}
\usepackage{comment}
\usepackage{mathrsfs}
\usepackage{xcolor}
\usepackage[utf8]{inputenc}
\usepackage[footnotesize]{caption}
\usepackage{bbold}
\usepackage{xfrac}
\usepackage{physics}
\usepackage{tikz}
\usepackage[compat=1.1.0]{tikz-feynhand}

\usepackage[lf]{Baskervaldx} 
\usepackage[bigdelims,vvarbb]{newtxmath} 
\usepackage[cal=boondoxo]{mathalfa} 

\newcommand{\Mpl}{M_{\text{Pl}}}

\def\thefootnote{\fnsymbol{footnote}}
\usepackage{multicol}
\definecolor{dark_red}{rgb}{0.7, 0., 0.}
\definecolor{light_pink}{rgb}{1,0.4,0.4}
\definecolor{light_blue}{rgb}{0.284602,0.317763,0.963947}
\definecolor{forestgreen}{HTML}{228B22}
\definecolor{ochre}{HTML}{CCAA2B}
\begin{document}
\hypersetup{pageanchor=false}
\begin{titlepage}

    \begin{center}

        \hfill KEK-TH-2623\\
        \hfill TU-1234

        \vskip 1in
            {\Huge \bfseries Thermal Wash-in Leptogenesis\\ via Heavy Higgs Decay\\}
        \vskip .8in
            {\Large Kyohei Mukaida$^{a}$, Hidenaga Watanabe$^{a}$, Masaki Yamada$^{b}$}

        \vskip .3in
        \begin{tabular}{ll}
            $^a$ & \!\!\!\!\!\emph{Theory Center, IPNS, KEK, 1-1 Oho, Tsukuba, Ibaraki 305-0801, Japan}                   \\
            $^a$ & \!\!\!\!\!\emph{Graduate University for Advanced Studies (Sokendai), }                                 \\[-.3em]
                 & \!\!\!\!\!\emph{1-1 Oho, Tsukuba, Ibaraki 305-0801, Japan}                                             \\
            $^b$ & \!\!\!\!\!\emph{FRIS, Tohoku University, Sendai, Miyagi 980-8578, Japan}                               \\
            $^b$ & \!\!\!\!\!\emph{Department of Physics, Tohoku University, Sendai, Miyagi 980-8578, Japan}              \\
        \end{tabular}

    \end{center}

    \vskip .6in

    \begin{abstract}
        \noindent
        We present a conceptually simple model to generate asymmetries that are not directly related to baryon nor lepton charges. The model employs a three-Higgs doublet framework, wherein the other two Higgs fields are significantly heavier than the Standard Model (SM) Higgs field. The decay of these heavier Higgs fields generates asymmetry for approximately conserved charges in the Standard Model at a high temperature. These asymmetries will be converted into baryon/lepton asymmetry through $B-L$ violating interactions associated with right-handed neutrinos via the wash-in mechanism.
    \end{abstract}

\end{titlepage}

\tableofcontents
\thispagestyle{empty}
\renewcommand{\thepage}{\arabic{page}}
\renewcommand{\thefootnote}{$\natural$\arabic{footnote}}
\setcounter{footnote}{0}
\newpage
\hypersetup{pageanchor=true}

\section{Introduction}
\label{sec:introduction}

The seesaw mechanism for neutrino mass generation is widely regarded as one of the most plausible theories extending beyond the Standard Model (SM)~\cite{Minkowski:1977sc,Yanagida:1979as,Yanagida:1980xy,Gell-Mann:1979vob,Mohapatra:1979ia}. This model naturally accounts for the small mass of left-handed neutrinos, as suggested by neutrino oscillation experiments. Additionally, it offers a straightforward method for generating baryon asymmetry through the decay of heavy right-handed neutrinos, a process known as vanilla thermal leptogenesis~\cite{Fukugita:1986hr} (see Refs.~\cite{Davidson:2008bu} for review). 

From another perspective, the existence of Majorana neutrinos or lepton-number violating interactions shifts the paradigm of baryogenesis. The $B-L$ violating interaction provides a scenario for baryogenesis originating from almost any approximate asymmetry above the energy scale of the $B-L$ violating interaction. This concept is analogous to how the discovery of the $B+L$ violating electroweak sphaleron process established a scenario for baryogenesis by generating $B-L$ asymmetry, commonly referred to as leptogenesis. In this process, baryon asymmetry is `washed-in' from lepton asymmetry via the electroweak sphaleron.
Similarly, lepton asymmetry can be `washed-in' from other asymmetries at high temperatures via $B-L$ violating interactions, a process we refer to as wash-in leptogenesis~\cite{Domcke:2020quw,Domcke:2022kfs,Schmitz:2023pfy}.

In the original paper, we analyzed the details of the wash-in leptogenesis without specifying a UV model to generate asymmetry above the $B-L$ violating scale. In this paper, we introduce a conceptually simple UV model to facilitate wash-in leptogenesis through the thermal history of the Universe following reheating. We propose a model incorporating three Higgs doublets, referred to as the three-Higgs doublet model (3HDM)~\cite{Keus:2013hya}, along with heavy right-handed neutrinos. Two of the Higgs doublets are assumed to be very heavy, and their decay generates asymmetry for charges that are approximately conserved in the SM at a high temperature. This process does not directly generate baryon or lepton asymmetry; however, the resulting asymmetry will be converted into baryon/lepton asymmetry by the wash-in mechanism, through $B-L$ violating interactions and the electroweak sphaleron process.
We find that the successful baryogenesis is realized for heavy Higgses with its mass being heavier than the intermediate scale and for the Majorana mass greater than $10^5 \, \mathrm{GeV}$.

\section{Wash-in Leptogenesis in a nutshell}
\label{sec:washin}

First, we briefly review the wash-in leptogenesis, focusing on the right-handed electron asymmetry.
We denote the mass of right-handed Majorana neutrinos as $M_i$ ($i=1,2,3$).
The decoupling temperature of $B-L$ violating interaction is represented by $T_{B-L}$.
We expect $T_{B-L} \simeq M_1$ as our main interest is $M_1 \ll 10^{12}\,\mathrm{GeV}$, where the $B-L$ violating interaction via the dimension five Weinberg operator is decoupled for $T \lesssim 10^{12}\,\mathrm{GeV}$.
For simplicity, we consider the case with $10^5 \,\mathrm{GeV} \lesssim T_{B-L} \lesssim 10^6 \,\mathrm{GeV}$
such that the $B-L$ violating interaction becomes decoupled in the regime where only the electron Yukawa interaction is decoupled in the SM sector.
Our scenario can also be applied for the case with a higher $T_{B-L}$, in which case the final result changes by a factor of $\mathcal{O}(1)$.

The right-handed electron asymmetry can be changed via the SM electron Yukawa interactions.
However, the wash-out rate of the right-handed electron asymmetry is $\gamma_{e_\text{R}} \simeq \kappa_{y_e} y_e^2 T$ with $\kappa_{y_e} \simeq 0.014$ being a numerical factor and $y_e$ being the electron Yukawa, which is smaller than the Hubble expansion rate at a temperature above $T_e = 8.7 \times 10^4\,\mathrm{GeV}$~\cite{Bodeker:2019ajh,Domcke:2020kcp}.\footnote{
    An order one factor depends on the definition of the decoupling condition.
}
Above this temperature, the right-handed electron charge is approximately conserved.

As we will see below, we consider the case in which a nonzero right-handed lepton asymmetry $q_e$ is generated by a heavy particle decay.
We denote the number density of particle and anti-particle of species $i$ as $n_i$ and $n_{\bar{i}}$, respectively.
Its asymmetry is denoted as $q_i \equiv n_i - n_{\bar{i}}$, which is related to the chemical potential as $\mu_i = (6/T^2) q_i/g_i$ with $g_i$ being multiplicity.
We particularly note that the total hypercharge conservation implies nonzero asymmetry in other particles to compensate the hypercharge asymmetry from nonzero $\mu_e$.
Moreover, the nonzero $\mu_e$ biases the other chemical potentials for SM species via the spectator processes such as~\cite{Domcke:2020kcp,Domcke:2020quw} (see also Refs.~\cite{Campbell:1992jd,Cline:1993vv,Cline:1993bd,Fukugita:2002hu}\footnote{
    The importance of non-equilibration for non-leptonic spectator processes was also pointed out in the flavored leptogenesis in Ref.~\cite{Garbrecht:2014kda}.
})
\begin{align}
 &\mu_{\mu} =  \mu_{\tau} =
 \frac{7}{481} \mu_e\,,
 \qquad
 \mu_{\ell_e} =
 -\frac{415}{962} \mu_e \,,
 \qquad
 \mu_{\ell_\mu} =  \mu_{\ell_\tau} =
 \frac{59}{962} \mu_e \,,
 \\
 &\mu_{u} =  \mu_{c} =  \mu_{t} =
 \frac{3}{37} \mu_e \,,
 \qquad
 \mu_{d} =  \mu_{s} = \mu_{b} =
 -\frac{6}{481} \mu_e \,,
 \qquad
  \mu_{Q_1} =  \mu_{Q_2} = \mu_{Q_3} =
 \frac{33}{962} \mu_e \,,
 \\
  &\mu_{H_1} =
 \frac{45}{962} \mu_e \,.
\end{align}
This particularly implies $B = L = 0$.
Let us emphasize that $\mu_e$ can be nonzero even if $B$ and $L$ charges vanish.

So far we do not include the effect of right-handed neutrinos and $B-L$ violating interactions.
Once the $B-L$ violating interactions are turned on,
$B-L$ (or $B$) can be induced from a nonzero bias $\mu_e$ through transport equations.
The physical reason behind this is that the neutrino Yukawa interaction can only wash-out the lepton asymmetry other than the right-handed electrons, which results in a nonzero total lepton asymmetry and hence $B-L$ asymmetry.
Particularly in the `strong wash-in' regime, where the right-handed neutrino Yukawa interactions are in chemical equilibrium,
we obtain~\cite{Domcke:2020quw}
\begin{equation}
 q_{B-L} = - \frac{3}{10} q_e \,.
\end{equation}
This is dubbed as wash-in leptogenesis, in the sense that the $B-L$ asymmetry is `washed-in' rather than `washed-out' under the chemical equilibrium associated with $B - L$ violating processes via the neutrino Yukawa interactions.
The resultant $B-L$ asymmetry remains as long as the $B-L$ violating interactions decouple earlier than the equilibration of the electron Yukawa interaction, \textit{i.e.},
\begin{equation} \label{eq:majoranalower}
    T_{B-L} \simeq M_1 > T_e \,,
\end{equation}
since otherwise all the asymmetries are washed-out if both the $B-L$ violating interactions and the electron Yukawa interactions are equilibrated.

The conversion factor from the original right-handed lepton asymmetry to the final baryon asymmetry is calculated as
\begin{equation}
 \frac{q_B}{s} \simeq 
  \frac{28}{79} \frac{q_{B-L}}{s} = 
 - \frac{42}{395} \frac{q_e}{s} \,,
\end{equation}
The observed baryon asymmetry of the Universe, $\left. q_B/s \right|_\text{(obs)} \simeq 8.6 \times 10^{-11}$, can be explained when
\begin{equation} \label{eq:etob}
 \left.\frac{q_e}{s} \right|_{T_{B-L}} \simeq - 8.1 \times 10^{-10} \quad \text{for} \quad T_{B-L} > T_e \,.
\end{equation}
In the next section, we provide a simple model to generate this asymmetry.

\section{Right-handed electron asymmetry via heavy Higgs decay}
\label{sec:heavyhiggs}

Now we consider a model that generates the right-handed electron asymmetry via a heavy particle decay in a thermal system in an expanding Universe.
The model is three-Higgs-doublet model (3HDM) with majorana neutrinos, where
we introduce three copies of Higgs fields that have Yukawa interactions to SM fermions in a similar way.
One of the Higgses is just the SM Higgs field
whereas the rest two are assumed to be much heavier than the electroweak scale and not to have a nonzero vacuum expectation value at the potential minimum.
The thermal decay of the heavy Higgs fields can produce some asymmetries for approximately conserved charges in the SM sector at a high temperature.

The Lagrangian density of our interest is given by
\begin{equation}
    \mathcal{L} = \mathcal{L}_{\text{SM}} + \mathcal{L}_{\text{RH}\nu} + \mathcal{L}_{HH} \,,
    \label{eq:lagrangian}
\end{equation}
where the first term describes the SM and the second term is the Lagrangian for Majorana right-handed neutrinos, \textit{i.e.,}
\begin{equation}
    \mathcal{L}_{\text{RH}\nu} = \bar{N}_i i \slashed{\partial} N_i - \frac{1}{2} M_i \bar{N}_i N_i
    - \qty( \bar{\ell}_{f} \tilde{H}_\text{SM} \lambda_{f i}^{\text{SM}} N_{i} + \text{H.c.}) \,,
\end{equation}
where the Majorana right-handed neutrino is $N_i$, and the left-handed lepton doublet is $\ell_f$, with the family indices being $i = 1,2,3$ and $f = e, \mu, \tau$ respectively.
The SM Higgs doublet is denoted by $H_\text{SM}$ and the conjugate is $\tilde H_\text{SM} = i \sigma_2 H_\text{SM}^\dag$.
The neutrino Yukawa coupling is represented by $\lambda_{f i}^{\text{SM}}$.
For simplicity, we assume that the Majorana masses are hierarchical, \textit{i.e.,} $M_1 \ll M_2 < M_3$.
One may consider a more complicated spectrum but the extension is straightforward.
The third term in Eq.~\eqref{eq:lagrangian} involves the heavy Higgses, which is given by
\begin{equation}
    \mathcal{L}_{HH} = \abs{D H_\alpha}^2 - M^2_{H_\alpha} \abs{H_\alpha}^2 - \qty( \bar{\ell}_{f} H_{\alpha} Y_{f f'}^{\alpha} e_{f'}+\bar{\ell}_{f} \Tilde{H}_{\alpha} \lambda_{f i}^{\alpha} N_{i} + \text{H.c.} ) \,,
\end{equation}
where the heavy Higgs doublets are denoted by $H_\alpha$ with $\alpha = 1,2$, and the corresponding charged lepton and neutrino Yukawa couplings are $Y_{f f'}^{\alpha}$ and $\lambda_{f i}^{\alpha}$ respectively.
The masses of the heavy higgeses are taken to be $M_{H_1} < M_{H_2}$.
We mainly consider the case where $M_{H_1}$ and $M_{H_2}$ are of the similar order.
Notice that these Yukawa interactions never break $L$ nor $B$ charges if we assign the lepton charge to the right-handed neutrinos.

To realize the thermal wash-in scenario, we consider the case in which the reheating temperature of the Universe is higher than $M_{H_{1}}$ so that the lighter heavy Higgs is thermally populated initially.
The successful wash-in is realized if the $B-L$ violating interaction decouples after the production of the right-handed electron asymmetry but before the equilibration of the electron Yukawa interactions.
In our case, this implies $T_e < T_{B-L} \simeq M_1  < M_{H_1} \lesssim M_{H_2} \ll M_2, M_3$.

It is known that the $CP$-violating out-of-equilibrium decay of a heavy particle with its mass $M$ can generate a charge asymmetry around its decoupling temperature $T_{\rm dec} \sim M$.
This is similar to the thermal GUT baryogenesis~\cite{Yoshimura:1978ex,Dimopoulos:1978kv,Toussaint:1978br,Weinberg:1979bt,Barr:1979ye} and thermal leptogenesis~\cite{Fukugita:1986hr}; however, we do not generate neither $B$ nor $L$ asymmetries via the heavy Higgs decay as the Yukawa interactions conserve $B$ and $L$.
In our case, the decay/inverse-decay channels of the heavy Higgs are
\begin{align}
    H_1 &\longleftrightarrow \bar e_{f'} + \ell_{f} \,, \qquad  H_1 \longleftrightarrow  \bar \ell_{f} + N_1 \,, \\
    \bar H_1 &\longleftrightarrow e_{f'} + \bar\ell_{f} \,, \qquad
    \bar H_1 \longleftrightarrow \ell_{f} + N_1 \,.
\end{align}
In the minimal scenario, only a single flavor in the lepton sector couples to the heavy Higgses, and the flavor of the right-handed lepton is assumed to be the same as the electron in the SM charged-lepton Yukawa.
Here one may intuitively understand the reason why we need both decay channels of $H_1 \to \bar e \ell$ and $H_1 \to \bar \ell N_1$ to have a nonvanishing right-handed lepton asymmetry for the single flavor case.
As we assume no primordial asymmetry in $H_1$ and $\bar H_1$, the excess in $e \bar \ell$ over $\bar e \ell$ after the heavy Higgs decay seemingly contradicts with the hypercharge neutrality.
However, the hypercharge neutrality is always maintained because the excess in $\bar \ell N_1$ over $\ell N_1$ compensates the positive hypercharge contribution of $e \bar \ell$ over $\bar e \ell$.
In fact, we do not have the $CP$-violating decay unless we have both Yukawa couplings of $Y$ and $\lambda$.
We also note that the right-handed electron asymmetry can be generated without the decay channel into the right-handed neutrinos if more than one lepton-flavors couple to the heavy Higgses.
Although the total right-handed lepton asymmetry vanishes in this case, the indivisual one can be nonzero, \textit{e.g.,} $q_e = - q_\mu$, while the flavored SM-lepton asymmetry is still vanishing.\footnote{
        This highlights the difference between the wash-in leptogenesis and the flavored leptogenesis~\cite{Abada:2006fw,Nardi:2006fx,Abada:2006ea,Blanchet:2006be,Pascoli:2006ie,DeSimone:2006nrs} because the wash-in leptogenesis never calls for the direct production of the flavored SM-lepton asymmetries, \textit{i.e.,} $q_{L_f} = 0$, but works only with the right-handed electron asymmetry from the decay of heavy Higgs field, $q_{e} \neq 0$.
        Later, the right-handed lepton asymmetry is reprocessed into the $B-L$ asymmetry via scatterings involving the right-handed neutrinos.
}
See Appendix~\ref{sec:appendix} for more details.
In the following, we only consider the single flavor case for simplicity.

\begin{figure}[t]
    \begin{minipage}[b]{0.45\linewidth}
    \centering
    \begin{tikzpicture}
    \begin{feynhand}
    \vertex[particle] (a) at (-0.3,0) {$H_1$}; \vertex (b) at (1,0);
    \vertex (c) at (2,0); \vertex (d) at (3.0,0);
    \vertex[particle] (e1) at (4.5,1.5) {$e_{f^\prime}$}; 
    \vertex[particle] (e2) at (4.5,-1.5) {$\ell_f$};
    \propag [chasca] (a) to (b);
    \propag[antfer] (b) to [edge label = $\ell_{f''}$] [in=90, out=90, looseness=1.7] (c);
    \propag[plain] (b) to [edge label' = $N_i$] [in=270, out=270, looseness=1.7] (c);
    \propag[chasca] (c) to [edge label = $H_2$](d);
    \propag[antfer] (d) to (e1);
    \propag[fer] (d) to (e2);
    \end{feynhand}
    \end{tikzpicture}
    \end{minipage}
    \begin{minipage}[b]{0.45\linewidth}
    \centering
    \begin{tikzpicture}
    \begin{feynhand}
    \vertex[particle] (a) at (-0.3,0) {$H_1$}; \vertex (b) at (1,0);
    \vertex (c1) at (2.8,0.8); \vertex (c2) at (2.8,-0.8);
    \vertex[particle] (d1) at (4.5,1.5) {$e_{f^\prime}$}; 
    \vertex[particle] (d2) at (4.5,-1.5) {$\ell_f$};
    \propag [chasca] (a) to (b);
    \propag[antfer] (b) to [edge label = $\ell_{f''}$]  (c1);
    \propag[plain] (b) to [edge label' = $N_i$] (c2);
    \propag[chasca] (c2) to [edge label' = $H_2$] (c1);
    \propag[antfer] (c1) to (d1);
    \propag[fer] (c2) to (d2);
    \end{feynhand}
    \end{tikzpicture}
    \end{minipage}
    \caption{Feynman diagrams responsible for the right-handed electron asymmetry. The interference of these one-loop diagrams with the tree-level diagram yields the required $CP$-violation.}
    \label{fig:feyn}
\end{figure}
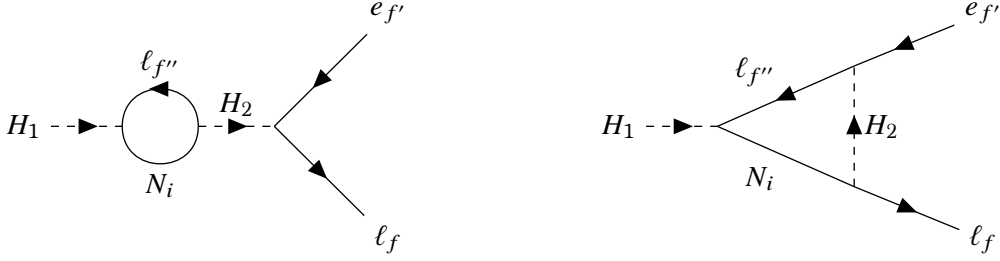

At a high temperature with $T > M_{H_1}$, we assume that $H_1$ and $\bar H_1$ is thermally populated and that there is no primordial asymmetries between them.
The initial abundance of $H_1$ and $\bar H_1$ is therefore
\begin{equation}
    \label{eq:abundance_H1}
    \frac{n_{H_1}}{s} = \frac{n_{\bar H_1}}{s} \simeq 2 \frac{n_\text{eq}}{s} = \frac{45\zeta (3)}{ \pi^4} \frac{1}{g_\ast} \,.
\end{equation}
To guarantee that the inverse decay does not wash-out the right-handed electron asymmetry, we require
\begin{equation} \label{eq:lowerbound}
    1 \gtrsim \left. \frac{\Gamma_{H_1 \to \bar e_1 \ell}}{H} \right|_{T = M_{H_1}}
    \simeq \qty(\frac{90}{\pi^2 g_\ast})^{1/2} \frac{ \qty(Y^{1\dag} Y^1)_{11}}{16 \pi } \frac{\Mpl}{M_{H_1}}
    \quad \longrightarrow \quad
    M_{H_1} \gtrsim 10^{12} \,\mathrm{GeV}\, \qty(\frac{Y}{0.01})^2 \,.
\end{equation}
Here we define
\begin{align}
    &\Gamma_{H_1 \to \bar e_f \ell} \equiv \sum_{f'} \Gamma_{H_1 \to \bar e_f \ell_{f'}} \,, \qquad
    \Gamma_{\bar H_1 \to e_f \bar\ell} \equiv \sum_{f'} \Gamma_{\bar H_1 \to e_f \bar\ell_{f'}} \,, 
    \\
    &\Gamma_{H_1 \to \bar e \ell} \equiv 
    \sum_{f} \Gamma_{H_1 \to \bar e_f \ell} \,,
    \qquad
    \Gamma_{\bar H_1 \to e \bar \ell} \equiv 
    \sum_f \Gamma_{\bar H_1 \to e_f \bar\ell} \,,
\end{align}
and $\Gamma_{H_1}$ as the total decay width for $H_1$.

Let us for a while neglect the wash-out and estimate the resultant right-handed electron asymmetry.
In the following discussion, we assume a hierarchy $M_1 \ll M_{H_1}$ and diagonal Yukawa couplings with respect to the SM lepton-flavor index, \textit{i.e.,} $Y^\alpha_{ff'} = Y^\alpha_{f} \delta_{ff'}$ and $\lambda^\alpha_{fi} = \lambda^\alpha_f \delta_{fi}$ with $f = e,\mu,\tau$, which is essentially reduced to the single flavor case.
The hierarchy of $M_1 \ll M_{H_1}$ is typically realized since the Higgs mass should be $M_{H_1} \gtrsim 10^{12} \,\mathrm{GeV}$ [Eq.~\eqref{eq:lowerbound}] while the Majorana mass can be as small as $M_1 \gtrsim 10^5 \,\mathrm{GeV}$ [Eq.~\eqref{eq:majoranalower}].
The right-handed electron asymmetry can be estimated as
\begin{align}
    \frac{q_{e_1}}{s} &= \frac{n_{e_1} - n_{\bar e_1}}{s} \simeq \frac{\Gamma_{\bar H_1 \to e_1 \bar \ell}}{\Gamma_{H_1}} \frac{n_{\bar H_1}}{s} - \frac{\Gamma_{H_1 \to \bar e_1 \ell}}{\Gamma_{H_1}} \frac{n_{H_1}}{s}
    \simeq
    - \frac{2 \Gamma_{H_1 \to \bar e \ell}^{(0)}}{\Gamma_{H_1}^{(0)} } \epsilon_{\bar e_1 \ell} \times \frac{45\zeta (3)}{ \pi^4} \frac{1}{g_\ast} \,,
\end{align}
where the superscript $(0)$ indicates the tree level contribution.
The asymmetry parameter of $H_1 \to \bar e_f \ell$ is defined by
\begin{equation}
    \epsilon_{\bar e_f \ell} \equiv\frac{\Gamma_{H_1 \to \bar e_f \ell} - \Gamma_{\bar H_1 \to e_f \bar \ell}}{\Gamma_{H_1 \to \bar e \ell} + \Gamma_{\bar H_1 \to e \bar \ell}} 
    \simeq \frac{\Gamma_{H_1 \to \bar e_f \ell} - \Gamma_{\bar H_1 \to e_f \bar \ell}}{2 \Gamma_{H_1 \to \bar e \ell}^{(0)}} ,
\end{equation}
As given in the appendix, the asymmetry parameter can be expressed as
\begin{equation}
    \epsilon_{\bar e_1 \ell} 
    \simeq \frac{1}{8 \pi} 
    \frac{\Im \qty( Y^{1}_{11} Y^{2\dag}_{11} \lambda^{1 }_{11} \lambda^{2\dag}_{11} )}{\tr \qty(Y^{1\dag} Y^1)}
    f(M_{H_2}/M_{H_1}) \,,
\end{equation}
where
\begin{equation}
    f(x) \equiv f_\text{self} (x) + f_\text{vertex} (x), \qquad
    f_\text{self} (x) \equiv \frac{1}{x^2 - 1}, \qquad f_\text{vertex} (x) \equiv 1 - x^2 \log \qty( 1 + 1/x^2 ) \,.
\end{equation}
The subscripts indicate that the respective contribution is originated from the self-energy diagram or the vertex diagram (see Fig.~\ref{fig:feyn}).
The tree-level decay widths are summarized as
\begin{equation}
    \Gamma_{H_1 \to \bar e \ell}^{(0)} \simeq \Gamma_{\bar H_1 \to e \bar\ell}^{(0)}
    \simeq \frac{1}{16 \pi} \tr \qty(Y^{1 \dag} Y^1) M_{H_1} \,, \qquad
    \Gamma_{H_1 \to \bar\ell N_1}^{(0)} \simeq \Gamma_{\bar H_1 \to \ell N_1}^{(0)}
    \simeq \frac{1}{16 \pi} \tr \qty(\lambda^{1 \dag} \lambda^1) M_{H_1} \,,
\end{equation}
for $M_1 \ll M_{H_1}$.
A complete forms of the asymmetry parameter with fully general flavor indices are given in the appendix.

Finally, by taking into account the wash-out factor or the inefficient production of $H_1$ collectively as $\kappa$, we obtain the following estimation for the right-handed electron asymmetry
\begin{align}
    \frac{q_{e}}{s} = \frac{q_{e_1}}{s} 
    &\simeq - \frac{\kappa}{4 \pi} \frac{\Im \qty( Y^{1}_{11} Y^{2\dag}_{11} \lambda^{1}_{11} \lambda^{2\dag}_{11} )}{\tr \qty(Y^{1 \dag} Y^1) + {\tr \qty(\lambda^{1 \dag} \lambda^1) }} f (M_{H_2}/M_{H_1}) \times \frac{45\zeta (3)}{ \pi^4} \frac{1}{g_\ast} \\
    & \sim - 8 \times 10^{-10} \delta_\text{CP}\, \qty( \frac{Y}{0.01} )^2 \,,
\end{align}
where we denote the $CP$-violating phase as $\delta_\text{CP}$, and we take $M_{H_2} = 2 M_{H_1}$, $Y^1 \sim Y^2 \sim \lambda^1 \sim \lambda^2$, and $\kappa \sim 0.1$ for simplicity.
One may readily confirm that the observed baryon asymmetry can be reproduced [see Eq.~\eqref{eq:etob}].

\section{Summary and discussion}
\label{sec:summary}

We have demonstrated that the baryon asymmetry of the Universe can be explained by the wash-in mechanism in a 3HDM with heavy right-handed neutrinos. Heavy Higgs fields, generated from the thermal plasma, undergo subsequent decay that produces certain asymmetries in the SM. The right-handed neutrinos are assumed to be much lighter than the heavy Higgs fields, ensuring that the $B-L$ violating process is efficient at a low temperature, which allows for the `wash-in' of lepton asymmetry. As the decay of the heavy Higgs does not directly generate baryon or lepton asymmetry, this represents a non-trivial UV-complete realization of the wash-in mechanism.

Our model bears similarities to the one considered in the context of Dirac leptogenesis~\cite{Dick:1999je}.\footnote{
    Essentially the same model in a different parameter region was also investigated in the context of flavored leptogenesis in Ref.~\cite{Garbrecht:2012qv}.
    There, the additional Higgs is assumed to be lighter than the right-handed neutrino, and the flavored lepton asymmetry is generated by the $CP$-violating decay of the right-handed neutrino.
    The additional Higgs provides an interference term with the SM Higgs, which leads to a new source of $CP$ violation for the right-handed neutrino decay.
    }
In that case, they considered very light right-handed neutrinos, and the $B-L$ violating process was not in thermal equilibrium. In contrast, our model involves relatively heavy right-handed neutrinos, with the $B-L$ violating process in thermal equilibrium in the early Universe. The particle contents are the same in both models, and the primary difference lies in the mass of the right-handed neutrinos. In this sense, the scenario we consider here can be seen as a complementary approach to Dirac leptogenesis, expanding the possibilities for a consistent cosmological model. This demonstrates that the concept of the wash-in mechanism provides alternative perspectives for existing scenarios of baryo/leptogenesis.

Although we have focused on the thermal production of heavy Higgses in this paper, nothing prevents us from extending this scenario to the non-thermal production, such as the inflaton decay, the primordial black hole evaporation, and so on~\cite{Schmitz:2023pfy}.
Also, the right-handed electron asymmetry is obviously not the single option to generate the baryon asymmetry, since the Higgs may couple to quarks and have more flavor-structures in its Yukawa interaction.
A comprehensive analysis of this model is left for the future study.

\section*{Acknowledgement}
K.\,M.\, was supported by JSPS KAKENHI Grant No.\ JP22K14044.
M.\,Y.\ was supported by MEXT Leading Initiative for Excellent Young Researchers, and by JSPS KAKENHI Grant No.\ JP20H05851 and JP23K13092.

\appendix

\section{Formula of CP-violating heavy Higgs decay}
\label{sec:appendix}

In this Appendix, we briefly summarize the results of the $CP$-violating heavy Higgs decay by keeping a fully general flavor structure in Yukawa interactions, and take into account the mass of the Majorana right-handed neutrinos.
The relevant diagrams are given in Fig.~\ref{fig:feyn}, where one finds the well-known self-energy and vertex contributions.

The relevant flavored asymmetry parameters are defined by
\begin{equation}
    \epsilon_{\bar e_{f'} \ell_{f}} \equiv \frac{\Gamma_{H_1 \to \bar e_{f'} \ell_{f}}- \Gamma_{\bar H_1 \to e_{f'} \ell_{f}}}{\Gamma_{H_1 \to \bar e \ell} + \Gamma_{\bar H_1 \to e \bar \ell}}, \qquad
    \epsilon_{\bar \ell_{f} N_i} \equiv \frac{\Gamma_{H_1 \to \bar \ell_{f} N_i}- \Gamma_{\bar H_1 \to \ell_{f} N_i}}{\Gamma_{H_1 \to \bar\ell N} + \Gamma_{\bar H_1 \to \ell N}}.
\end{equation}
Owing to the $CPT$ theorem, the following equality of the forward scattering amplitudes must hold; $\mathcal M_{H_1 \to H_1} = \mathcal M_{\bar H_1 \to \bar H_1}$.
Hence, the total decay width of $H_1$ and $\bar H_1$ is the same
\begin{equation}
    \sum_{f, f'}\Gamma_{H_1 \to \bar e_{f'} \ell_{f}} + \sum_{f,i} \Gamma_{H_1 \to \bar \ell_{f} N_i} = \sum_{f, f'}\Gamma_{\bar H_1 \to e_{f'} \bar \ell_{f}} + \sum_{f,i} \Gamma_{\bar H_1 \to \ell_{f} N_i},
\end{equation}
which implies
\begin{equation}
    \label{eq:CPT_rel}
    \sum_{f,f'}\epsilon_{\bar e_{f'} \ell_{f}} \qty( \Gamma_{H_1 \to \bar e \ell} + \Gamma_{\bar H_1 \to e \bar \ell} ) = - \sum_{f,i}\epsilon_{\bar\ell_f N_i} \qty( \Gamma_{H_1 \to \bar \ell N_1} + \Gamma_{\bar H_1 \to \ell N_1} ).
\end{equation}
We will use this equation as a non-trivial sanity check of our analytical calculations.

In Appendix~\ref{sec:appB} and \ref{sec:appendixC}, we provide detailed calculations of $CP$-violating heavy Higgs decay into SM leptons and right-handed neutrinos for the case with nonzero $M_i$, respectively. 
Here we summarize the results.
The asymmetry parameter for $H_1 \to \bar e_{f'} \ell_{f}$ is given by
\begin{equation} \label{eq:self_e}
    \epsilon^\text{(self)}_{\bar e_{f'} \ell_{f}} =
    \frac{1}{8 \pi} \frac{ \Im\left[ Y_{ff'}^{1} (Y^{2 \dagger} )_{f'f} \qty( \lambda^{2 \dag} \lambda^{1} )_{ii}\right]}{\tr \qty( Y^{1\dag} Y^{1} )} f_\text{self} (M_{H_2}/M_{H_1}, M_i/M_{H_1}),
\end{equation}
for the self-energy contribution, and
\begin{equation}\label{eq:vertex_e}
    \epsilon^\text{(vertex)}_{\bar e_{f'} \ell_{f}}
    =\frac{1}{8 \pi} \frac{ \operatorname{Im}\left[ Y_{f f'}^{1} \qty(Y^{2\dag}  \lambda^{1 })_{f'i}\,  \lambda^{2 \dag}_{i f}  \right]}{ \tr \qty( Y^{1\dag} Y^{1} )} f_\text{vertex} (M_{H_2} / M_{H_1} , M_i/M_{H_1}),
\end{equation}
for the vertex contribution.
We defined the following functions
\begin{equation}
    f_\text{self} (x,y) \equiv \frac{(1-y^2)^2}{x^2 - 1}, \qquad
    f_\text{vertex} (x,y) \equiv 1 - y^2 - x^2 \log \qty( 1 + (1-y^2)/x^2 ).
\end{equation}
Note that they are nonzero if $x>1$ and $0 \le y<1$. 
Here we neglect the contributions from SM lepton loops in the self-energy diagram that will be commented shortly.
One may readily confirm the formula given in the main text by taking $M_1 / M_{H_1} \ll 1$ and $M_{2,3} / M_{H_1} \gg 1$.

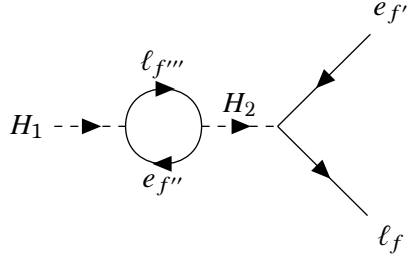
\begin{figure}[t]
    \centering
    \begin{tikzpicture}
    \begin{feynhand}
    \vertex[particle] (a) at (-0.3,0) {$H_1$}; \vertex (b) at (1,0);
    \vertex (c) at (2,0); \vertex (d) at (3.0,0);
    \vertex[particle] (e1) at (4.5,1.5) {$e_{f^\prime}$}; 
    \vertex[particle] (e2) at (4.5,-1.5) {$\ell_f$};
    \propag [chasca] (a) to (b);
    \propag[fer] (b) to [edge label = $\ell_{f'''}$] [in=90, out=90, looseness=1.7] (c);
    \propag[antfer] (b) to [edge label' = $e_{f''}$] [in=270, out=270, looseness=1.7] (c);
    \propag[chasca] (c) to [edge label = $H_2$](d);
    \propag[antfer] (d) to (e1);
    \propag[fer] (d) to (e2);
    \end{feynhand}
    \end{tikzpicture}
    
    \caption{Feynman diagrams responsible for the right-handed electron asymmetry, which becomes nonzero if the Yukawa involves more than one flavors. While it does not generate the total right-handed lepton asymmetry, individual right-handed lepton asymmetries can be generated, including the right-handed electron.}
    \label{fig:feyn2}
\end{figure}

We also note that the right-handed electron asymmetry can also be generated via the diagram without the right-handed neutrino in the loop. 
The relevant diagram is given in Fig.~\ref{fig:feyn2}, which is obtained by replacing $\ell_{f''}$ and $N_i$ with $e_{f''}$ and $\ell_{f'''}$, respectively in the self-energy diagram.
Note that the vertex diagram vanishes in this case.
This gives an additional contribution to the asymmetry parameter of Eq.~\eqref{eq:self_e} with the replacement of $\lambda^\alpha$ by $Y^{\alpha *}$:
\begin{equation}
    \label{eq:selfe_onlylep}
    \epsilon^\text{(self')}_{\bar e_{f'} \ell_{f}} = 
    \frac{1}{8 \pi} \frac{ \Im\left[ Y_{ff'}^{1} (Y^{2 \dagger} )_{f'f} \tr \qty( Y^{1\dag} Y^{2} )\right]}{\tr \qty( Y^{1\dag} Y^{1} )} f_\text{self} (M_{H_2}/M_{H_1}).
\end{equation}
Although the total asymmetry, $\sum_{f f'} \epsilon^\text{(self')}_{\bar{e}_{f'} \ell_f}$, vanishes, the individual one, $\sum_f \epsilon^\text{(self')}_{\bar{e}_{f'} \ell_f}$, does not in general for the case with three flavors.
Note that the flavored SM-lepton charges are vanishing as they are conserved in this case.

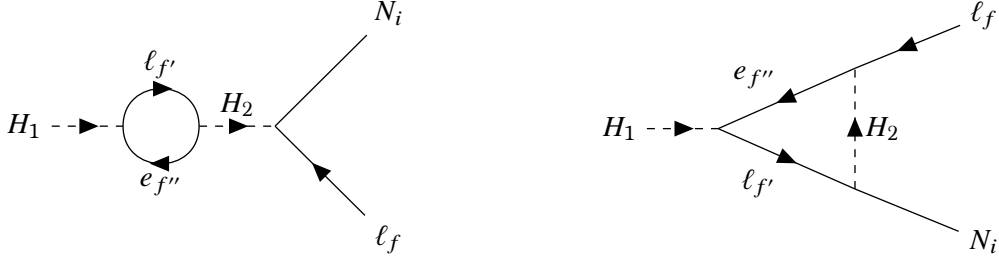
\begin{figure}[t]
    \begin{minipage}[b]{0.45\linewidth}
    \centering
    \begin{tikzpicture}
    \begin{feynhand}
    \vertex[particle] (a) at (-0.3,0) {$H_1$}; \vertex (b) at (1,0);
    \vertex (c) at (2,0); \vertex (d) at (3.0,0);
    \vertex[particle] (e1) at (4.5,1.5) {$N_i$}; 
    \vertex[particle] (e2) at (4.5,-1.5) {$\ell_f$};
    \propag [chasca] (a) to (b);
    \propag[fer] (b) to [edge label = $\ell_{f'}$] [in=90, out=90, looseness=1.7] (c);
    \propag[antfer] (b) to [edge label' = $e_{f''}$] [in=270, out=270, looseness=1.7] (c);
    \propag[chasca] (c) to [edge label = $H_2$](d);
    \propag[plain] (d) to (e1);
    \propag[antfer] (d) to (e2);
    \end{feynhand}
    \end{tikzpicture}
    \end{minipage}
    \begin{minipage}[b]{0.45\linewidth}
    \centering
    \begin{tikzpicture}
    \begin{feynhand}
    \vertex[particle] (a) at (-0.3,0) {$H_1$}; \vertex (b) at (1,0);
    \vertex (c1) at (2.8,0.8); \vertex (c2) at (2.8,-0.8);
    \vertex[particle] (d1) at (4.5,1.5) {$\ell_{f}$}; 
    \vertex[particle] (d2) at (4.5,-1.5) {$N_i$};
    \propag [chasca] (a) to (b);
    \propag[antfer] (b) to [edge label = $e_{f''}$]  (c1);
    \propag[fer] (b) to [edge label' = $\ell_{f'}$] (c2);
    \propag[chasca] (c2) to [edge label' = $H_2$] (c1);
    \propag[antfer] (c1) to (d1);
    \propag[plain] (c2) to (d2);
    \end{feynhand}
    \end{tikzpicture}
    \end{minipage}
    \caption{Feynman diagrams responsible for compensating the hypercharge excess in Fig.~\ref{fig:feyn}. When the total right-handed lepton asymmetry is generated by contributions in Fig.~\ref{fig:feyn}, these processes completely cancel out the apparent hypercharge excess as ensured by the hypercharge neutrality.}
    \label{fig:feyn_N}
\end{figure}

The asymmetry parameter for $H_1 \to \bar \ell_{f} N_i$ is given by 
\begin{equation} \label{eq:self_N}
    \epsilon^\text{(self)}_{\bar \ell_{f} N_i} = 
    - \frac{1}{8 \pi} \frac{ \Im\left[ \lambda_{fi}^{1} (\lambda^{2 \dagger} )_{if} \tr \qty( Y^{1} Y^{2 \dag} )\right]}{\tr \qty( \lambda^{1\dag} \lambda^{1} )} f_\text{self} (M_{H_2}/M_{H_1}, 0),
\end{equation}
for the self-energy contribution, and
\begin{equation}\label{eq:vertex_N}
    \epsilon^\text{(vertex)}_{\bar \ell_{f} N_i} =- \frac{1}{8 \pi} \frac{ \operatorname{Im}\left[ \lambda_{f i}^{1} \qty(\lambda^{2\dag}  Y^{1 } Y^{2 \dag} )_{if}   \right]}{ \tr \qty( \lambda^{1\dag} \lambda^{1} )} \frac{f_\text{vertex} (M_{H_2} / M_{H_1}, M_i / M_{H_1})}{1-M_i^2 / M_{H_1}^2} ,
\end{equation}
for the vertex contribution, where, for simplicity, we neglect the contributions from right-handed neutrino loops, since they vanish after the summation over the flavor indices as in the case with Eq.~\eqref{eq:selfe_onlylep}.

The tree level partial decay width of $H_1$ is given by
\begin{equation}
    \Gamma_{H_1 \to \bar e_{f'} \ell_{f} }^{(0)} = \Gamma_{\bar H_1 \to e_{f'} \bar \ell_{f} }^{(0)}
    = \frac{\abs{Y_{f f'}}^2}{16\pi} M_{H_1}, \qquad
    \Gamma_{H_1 \to \bar \ell_{f} N_i}^{(0)} = \Gamma_{\bar H_1 \to \ell_{f} N_i}^{(0)} 
    = \frac{\abs{\lambda_{f i}}^2}{16\pi} \frac{\qty( M_{H_1}^2 - M_i^2)^2}{M_{H_1}^3}.
\end{equation}
Noting that we have $\Gamma_{H_1 \to \bar e_{f'} \ell_f} + \Gamma_{\bar H_1 \to e_{f'} \bar \ell_f} = 2 \Gamma_{H_1 \to \bar e_{f'} \ell_f}^{(0)} $ and $\Gamma_{H_1 \to \bar \ell_f N_i} + \Gamma_{\bar H_1 \to \ell_f N_i} = 2 \Gamma_{H_1 \to \bar \bar \ell_f N_i}^{(0)} $ at this order, one may confirm Eq.~\eqref{eq:CPT_rel} explicitly from Eqs.~\eqref{eq:self_e}, \eqref{eq:vertex_e}, \eqref{eq:self_N}, and \eqref{eq:vertex_N}.

\section{Detailed calculations of CP-violating heavy Higgs decay into SM leptons}
\label{sec:appB}

In this Appendix, we show the detailed calculations for the asymmetry of right-handed charged leptons generated from the heavy-Higgs-boson decay, including the dependence on the right-handed neutrino masses, $M_i$. We denote the amplitude for the heavy-particle decay, $H_{1} \to \bar e_{f'} + l_{f} $, such as
\begin{equation}
    \mathcal{M}_{H_{1} \to \bar e_{f'} l_{f}}=(c_{0})_{ff'} \mathcal{A}_{0}+(c_{1})_{ff'} \mathcal{A}_{1}\,,
\end{equation}
where the subscripts $0$ and $1$ denote the quantities under the tree level and the one-loop level, respectively. We factorize the dependence on coupling constants in the amplitude as $c_0$ and $c_1$, and denote the rest parts as $\mathcal{A}$.
We also denote
the amplitude for CP-conjugated process as $\mathcal{M}_{\bar{L}}$.
It can be represented as
\begin{equation}
\mathcal{M}_{\bar H_{1} \to e_{f'} \bar l_{f}}= (c_{0}^{*})_{ff'} \mathcal{A}_{0} + (c_{1}^{*})_{ff'} \mathcal{A}_{1} \,.
\end{equation}
Then the asymmetry is calculated as 
\begin{align}
\epsilon_{\bar e_{f'} \ell_{f}}
&= 
\frac{\left|\mathcal{M}_{H_{1} \to \bar e_{f'} l_{f}}\right|^{2}-\left|\mathcal{M}_{\bar H_{1} \to e_{f'} \bar l_{f}}\right|^{2}}{\sum_{ff'} \left|\mathcal{M}_{H_{1} \to \bar e_{f'} l_{f}}\right|^{2}+ \sum_{ff'} \left|\mathcal{M}_{\bar H_{1} \to e_{f'} \bar l_{f}}\right|^{2}}
\\
&\simeq \frac{-2 \operatorname{Im}\left[ (c_{0}^{*})_{ff'} (c_{1})_{ff'} \right] \operatorname{Im}\left(\mathcal{A}_{0}^{*} \mathcal{A}_{1}\right)}{\sum_{ff'} \left| (c_{0})_{ff'} \mathcal{A}_{0}\right|^{2}}\,.
\label{asymmetry}
\end{align}
Thus the asymmetry is generated from the interference between the tree and the one-loop diagrams.

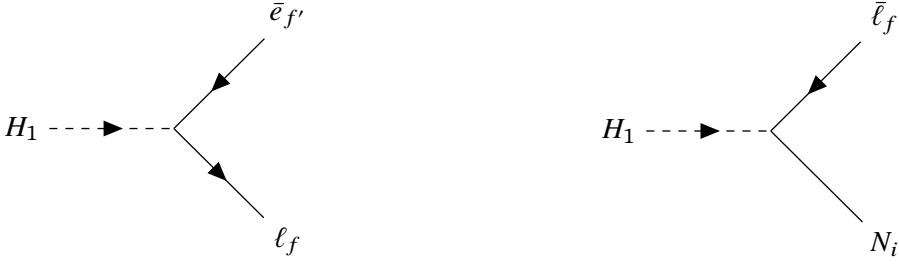
\begin{figure}[t]
    \begin{minipage}[b]{0.45\linewidth}
    \centering
    \begin{tikzpicture}
    \begin{feynhand}
    \vertex[particle] (a) at (1.,0) {$H_1$};
    \vertex (d) at (3.0,0);
    \vertex[particle] (e1) at (4.5,1.5) {$\bar e_{f^\prime}$}; 
    \vertex[particle] (e2) at (4.5,-1.5) {$\ell_f$};
    \propag [chasca] (a) to (d);
    \propag[antfer] (d) to (e1);
    \propag[fer] (d) to (e2);
    \end{feynhand}
    \end{tikzpicture}
    \end{minipage}
    \begin{minipage}[b]{0.45\linewidth}
    \centering
    \begin{tikzpicture}
    \begin{feynhand}
        \vertex[particle] (a) at (1.,0) {$H_1$};
        \vertex (d) at (3.0,0);
        \vertex[particle] (e1) at (4.5,1.5) {$\bar \ell_{f}$}; 
        \vertex[particle] (e2) at (4.5,-1.5) {$N_i$};
        \propag [chasca] (a) to (d);
        \propag[antfer] (d) to (e1);
        \propag[plain] (d) to (e2);
    \end{feynhand}
    \end{tikzpicture}
    \end{minipage}
    \caption{Feynman diagrams responsible for the tree level decay of $H_1$ into $\bar e_{f'} + \ell_{f}$ and $H_1$ into $\bar \ell_{f} + N_{i}$.}
    \label{fig:tree_e}
\end{figure}

The tree diagram is shown in Fig.~\ref{fig:tree_e}, whose corresponding contribution is given by
\begin{align}
\sum_{ff'} \left|(c_{0})_{ff'} \mathcal{A}_{0}\right|^{2} 
& =\sum_{ff'} \sum_{s,r} \sum_b \bar{u}^s(p+q) (-i) P_{R} Y_{ff'}^{1} \delta_{a b} v^r(-q) \bar{v}^r(-q)i  \delta_{a b} Y_{ff'}^{1 *} P_{L} u^s(p+q) \\
& = \sum_{ff'} \left|Y_{ff'}^{1}\right|^{2} M_{H_{1}}^{2}\,.
\end{align}
where we take the summation for the $\mathrm{SU}(2)$ index $b$ and spin indices $s,r$ for the final states, and we use $2 p_\mu q^\mu = - M_{H_{1}}^2$ and $P_L = (1-\gamma_5)/2 \to 1/2$.
The tree level decay rate is then given by 
\begin{equation}
\Gamma_{H_1 \to \bar e_{f'} \ell_{f} }^{(0)} 
= \frac{\abs{Y_{f f'}}^2}{16\pi} M_{H_1},  
\end{equation}

The one-loop diagrams include the contributions from the self-energy correction and that from the vertex correction. We decompose 
\begin{align}
&(c_1)_{ff'} = (c_1^{\rm (self)})_{ff'} + (c_1^{\rm (vertex)})_{ff'}\,,
\\
&\mathcal{A}_{1} = \mathcal{A}_{1}^{(\rm self)} + \mathcal{A}_{1}^{(\rm vertex)}\,,
\\
&\epsilon_{\bar e_{f'} \ell_{f}} = \epsilon_{\bar e_{f'} \ell_{f}}^{(\rm self)} + \epsilon_{\bar e_{f'} \ell_{f}}^{(\rm vertex)}\,,
\end{align}
accordingly.

\subsection{Self-energy correction}

The left diagram in Fig.~\ref{fig:feyn} 
represent the self-energy correction. 
The amplitude is given by 
\begin{align}
(c_1^{\rm (self)})_{ff'} \mathcal{A}_1^{\rm (self)} 
&= \sum_{i,j} \sum_{c,d} \bar{u}(p+q) (-i) P_{R} Y_{ff'}^{2} \delta_{b c} v(-q) \frac{i}{p^{2}-M_{H_{2}}^{2}} 
\\
&\quad \times(-1) \int \frac{\dd^{4} k}{(2 \pi)^{4}}\mathrm{tr}\left[(-i) \varepsilon_{d c} \lambda_{j i}^{2} P_{R} \frac{i(\slashed k+\slashed p)+M_{i}}{(k+p)^{2}-M_{i}^{2}}  (-i) \varepsilon_{d a} \lambda_{j i}^{1 *} P_{L} \frac{i \slashed k}{k^{2}}\right]
\nonumber\\
& = -2 \delta_{a b} \sum_{i,j} Y_{ff'}^{2} \lambda_{ji}^{2} \lambda_{ji}^{1 *} \frac{i  }{p^{2}-M_{H_{2}}^{2}}\left( A+ p_{\mu} B^{\mu}\right) \bar{u}(p+q) P_{R} v(-q) \,,
\end{align}
with the total anti-symmetric tensor of $\mathrm{SU}(2)$ being $\varepsilon_{ab}$.
Here we define 
\begin{align}
    &A \equiv -i \int \frac{\dd^4 k}{(2\pi)^4} \frac{1}{(k+p)^2-M_i^2} 
        \label{eq:A}\\
    &B^{\mu} \equiv -i \int \frac{\dd^4 k}{(2\pi)^4} \frac{k^{\mu}}{\qty[(k+p)^2-M_i^2 ]  k^2} \,.
    \label{eq:B}
\end{align}

We note that the interference between the tree and self-energy is written as 
\begin{align}
\mathcal{A}_{0}^{*} \mathcal{A}^{\rm (self)}_{1}
&= \sum_{s,r} \sum_b \bar{v}^r(-q) \delta_{a b} P_{L} i  u^s(p+q) \frac{-2 i \delta_{a b}}{p^{2}-M_{H_{2}}^{2}}\left(A+p_{\mu} B^{\mu}\right) \bar{u}^s(p+q) P_{R} v^r(-q) \\
&  =- \frac{2 M_{H_{1}}^{2} }{M_{H_{2}}^{2}-M_{H_1}^2}\left(A+p_{\mu} B^{\mu}\right)  \,. \label{treeself}
\end{align}
This implies that 
we need a nonzero imaginary part for $A$ or $ p_{\mu} B^{\mu}$ 
to generate the asymmetry. 
We note that the quadratic divergence does not give an imaginary part contribution. In particular, 
the finite part of $A$ is real 
after the subtraction from the wavefunction renormalization factor.

We also note a useful formula: 
\begin{align}
&\Im \qty[ \int_{0}^{1} \dd x \log \qty( f(x) -i \epsilon) ]
\nonumber\\
&= \Im \qty[ \int_{0}^{\alpha} \dd x \qty(\log [ - f(x)] -i\pi) ]
\nonumber\\
&= - \pi \alpha  \,,
\label{eq:formula}
\end{align}
where $f(x) < 0$ for $0 < x < \alpha$ and $f(x) > 0$ for $\alpha < x < 1$. 
Similarly, 
\begin{align}
&\Im \qty[ \int_{0}^{1} \dd x \, g(x) \log \qty( f(x) -i \epsilon) ]
\nonumber\\
&= -\pi \int_{\beta}^1 \dd x \, g(x)  \,,
\label{eq:formula2}
\end{align}
where $f(x) > 0$ for $0 < x < \beta$ and $f(x) < 0$ for $\beta < x < 1$.

We thus need to calculate $ p_{\mu} B^{\mu}$. 
It is rewritten as 
\begin{align}
p_{\mu} B^{\mu} & = - i \int \frac{\dd^{4} k}{(2 \pi)^{4}} \frac{p \cdot k}{\qty[(k+p)^{2}-M_i^{2} ]  k^{2}} \\
& =\frac{- i }{2} \int \frac{\dd^{4} k}{(2 \pi)^{4}} \frac{\qty[(k+p)^{2}-M_i^{2} ] -k^{2}-p^{2}+M_i^{2}}{\qty[(k+p)^{2}-M_i^{2} ]  k^{2}}  \\
& =\frac{- i }{2} \int \frac{\dd^{4} k}{(2 \pi)^{4}} \frac{1}{k^{2}}-\frac{- i }{2} \int \frac{\dd^{4} k}{(2 \pi)^{4}}\frac{1}{(k+p)^{2}-M_i^{2}} 
+\frac{1}{2}\left(M_{H_{1}}^{2}-M_i^{2}\right) \int \frac{\dd^{4} k}{(2 \pi)^{4}} \frac{i}{\qty[(k+p)^{2}-M_i^{2} ]  k^{2}} \,,
\end{align}
where we use $p^2 = M_{H_{1}}^2$. 
Again, 
the first and second terms are real after the subtraction of wavefunction renormalizaiton factor. 
The finite part of the integral in the third term is
\begin{align}
    \frac{1}{16 \pi^{2}} \int_{0}^{1} \dd x \log \qty{ M_{H_{1}}^{2} x \qty[x- \qty(1-\frac{M_i^2}{M_{H_{1}}^2}) ]-i \epsilon } \,.
    \label{eq:self-pB}
\end{align}
Using \eqref{eq:formula} with $\alpha =1-M_i^2/M_{H_{1}}^2$, we obtain 
\begin{align}
\operatorname{Im} \left(p_{\mu} B^{\mu}\right) =  - \frac{1}{32 \pi} \frac{\left(M_{H_{1}}^{2}-M_i^{2}\right)^{2}}{M_{H_{1}}^{2}} \,, 
\label{eq:bB}
\end{align}
for $M_i < M_{H_{1}}$.  
This leads to 
\begin{align}
\operatorname{Im}\left(\mathcal{A}_{0}^{*} \mathcal{A}^{\rm (self)}_{1}\right) = \frac{1}{16 \pi} \frac{\left(M_{H_{1}}^{2}-M_i^{2}\right)^{2}}{M_{H_{2}}^{2}-M_{H_{1}}^{2}} \,.
\label{eq:Imself}
\end{align}
Note that 
\eqref{eq:self-pB} is real and the self-energy does not lead to asymmetry 
for the case with $M_i > M_{H_{1}}$ or $\alpha < 0$.

From \eqref{asymmetry} and \eqref{eq:Imself}, 
the asymmetry from the self-energy correction is given by 
\begin{align}
\epsilon^{\rm (self)}_{\bar e_{f'} \ell_{f}} = -\frac{1}{8 \pi} \frac{\sum_{ij} \operatorname{Im}\left(Y_{ff'}^{2} Y_{ff'}^{1 *} \lambda_{j i}^{2} \lambda_{j i}^{1 *}\right)}{\sum_{ff'} \left|Y_{ff'}^{1}\right|^{2}} \frac{1}{M_{H_{1}}^{2}} \frac{\left(M_{H_{1}}^{2}-M_i^{2}\right)^{2}}{M_{H_{2}}^{2}-M_{H_{1}}^{2}} \,,
\end{align}
for $M_i<M_{H_{1}}$.

\subsection{Vertex correction}

The right diagram in Fig.~\ref{fig:feyn} 
represent the vertex correction. 
The amplitude is given by 
\begin{align}
(c_1^{\rm (vertex)})_{ff'} \mathcal{A}_1^{\rm (vertex)}
&= \sum_{i,j} \sum_{c,d} \bar{u}(p+q)(-i)\varepsilon_{bc}\lambda_{fi}^{2} P_R \int \frac{\dd^4 k}{(2\pi)^4} \frac{i\qty[(\slashed p + \slashed q + \slashed k) + M_i  ] }{(p+q+k)^2-M_i^2} 
\nonumber
\\
&\qquad \times  (-i) \varepsilon_{da} \lambda_{f''i}^{1 *}  P_L \frac{i(\slashed q + \slashed k)}{(q+k)^2} (-i) P_R Y_{f'' f'}^{2} \delta_{dc} \frac{i}{k^{2}-M_{H_{2}}^{2}} v(-q)
\nonumber\\
&= \delta_{a b} \sum_{i,j} \lambda_{f i}^{2} \lambda_{f'' i}^{1 *} Y_{f'' f'}^{2} (-i) D \bar{u}(p+q) P_{R}v(-q) \,,
\end{align}
where we use the equation of motion for outgoing fermions and $\slashed k \slashed k = k^2$ to simplify the equation.
Here, $D$ is defined by 
\begin{align}
    D &\equiv  -i \int \frac{\dd^{4} k}{(2 \pi)^{4}} \frac{k^2}{\qty[(p+q+k)^{2}-M_i^{2} ] (q+k)^{2}\left(k^{2}-M_{H_{2}}^{2}\right)}
      \label{Ddef} \,, 
    \\
      &= -i \int \frac{\dd^{4} k}{(2 \pi)^{4}} \int_{0}^1 \dd x \int_{0}^1 \dd y \int_{0}^1 \dd z \frac{2 \delta(1-x-y-z)k^2}{\left\{ \qty[(p+q+k)^{2}-M_i^{2} ] x+(q+k)^{2} y +(k^{2}-M_{H_{2}}^{2})z \right\}^3}\\
      &= -i \int \frac{\dd^{4} l}{(2 \pi)^{4}} \int_0^1 \dd y \int_0^{1-y} \dd x \frac{2\left( l^2-M^2_{H_{1}}xy\right)}{\left(l^2-\Delta^{\prime} \right)^3} 
      \\
      &\equiv  D_1 + D_2 
      \label{D} \,, 
\end{align}
with 
\begin{align}
    &\Delta^{\prime} \equiv (M^2_i-yM_{H_{1}}^2-M_{H_{2}}^2)x-M^2_{H_{2}}y+M_{H_{2}}^2 \,. 
    \\
    &D_1 \equiv -i \int \frac{\dd^{4} l}{(2 \pi)^{4}} \int_0^1 \dd y \int_0^{1-y} \dd x \frac{2 l^2}{\left(l^2-\Delta^{\prime} \right)^3} 
    \\
    &D_2 \equiv -i \int \frac{\dd^{4} l}{(2 \pi)^{4}} \int_0^1 \dd y \int_0^{1-y} \dd x \frac{-2 M^2_{H_{1}}xy}{\left(l^2-\Delta^{\prime} \right)^3} \,.
\end{align}
where we use $(p+q)^2 = 0$ and $2 p \cdot q = -M_{H_1}^2$.

We note that  
the interference between tree and vertex gives 
\begin{align}
\mathcal{A}_{0}^{*} \mathcal{A}^{\rm (vertex)}_{1}
& =\sum_{s,r} \sum_b \bar{v}^r(-q) \delta_{a b} P_{L} i u^s(p+q)\delta_{a b} (-i) D \bar{u}^s(p+q) P_{R}v^r(-q) \\
& = M_{H_{1}}^2 D \,.
\label{treevertex}
\end{align}
Thus we need a nonzero imaginary part of $D$ to generate asymmetry.

Using the $\overline{\text{MS}}$ renormalization scheme, we obtain a finite part of $D_1$ as 
\begin{align}
    D_1 
    &\to  -\frac{1}{8\pi^2} \int_0^1 \dd y \int_0^{1-y} \dd x  \left(\log \Delta^{\prime} + \frac{1}{2}\right)
    \\
& = 
- \frac{1}{8 \pi^{2}} \int_{0}^{1} \dd y\left[\frac{ y M_{H_{1}}^{2}-M_i^{2}}{y M_{H_{1}}^{2}+M_{H_{2}}^{2}-M_i^{2}} (1-y) \log \qty[(M_i^{2}-y M_{H_{1}}^{2})(1-y) -i \epsilon ] \right. \nonumber\\
& \left.\quad + \frac{M_{H_{2}}^{2}(1-y)}{y M_{H_{1}}^{2} +M_{H_{2}}^{2}-M_i^{2}} \log \qty[M_{H_{2}}^{2}(1-y) ] - (1-y) \right] \,.
\label{second of first of D}
\end{align}
Using \eqref{eq:formula2} with 
$\beta = M_i^2/M_{H_{1}}^{2}$, 
we obtain 
\begin{align}
\Im D_1 &= 
\frac{1}{8 \pi} \int_{\beta}^{1} \dd y \frac{ y M_{H_{1}}^{2} -M_i^{2}}{y M_{H_{1}}^{2} +M_{H_{2}}^{2}-M_i^{2}} (1-y) \\
&= \frac{1}{16 \pi M_{H_1}^4} 
\qty[
\qty( M_{H_1}^2 - M_i^2) \qty( M_{H_1}^2 + 2 M_{H_2}^2 - M_i^2 )
+ 2 M_{H_2}^2 \qty( M_{H_1}^2 + M_{H_2}^2 - M_i^2 ) 
\log \qty( \frac{M_{H_2}^2}{M_{H_1}^2 + M_{H_2}^2 - M_i^2} )
]\,.
\label{eq:ImD1}
\end{align}
for $\beta < 1$. 
Note that the imaginary part vanishes if $\beta > 1$, 
\textit{i.e.}, $M_i > M_{H_1}$. 

Next we calculate $D_2$. By performing the $l$ integral, we obtain 
\begin{align} \label{last D}
D_2 &= 
\frac{ M_{H_{1}}^{2}}{16 \pi^{2}} \int_{0}^{1} \dd y  \int_0^{1-y} \dd x \frac{x y}{\qty(M_i^{2}-y M_{H_1}^2 - M_{H_{2}}^{2}) x + M_{H_2}^2(1-y)}
\\
&=\frac{ M_{H_{1}}^{2}}{16 \pi^{2}} \int_{0}^{1} \dd y \frac{y}{M_i^{2}-y M_{H_{1}}^{2} - M_{H_2}^2} \nonumber \\
& \qquad \qquad \times \qty[ 
(1-y) - \frac{M_{H_2}^2(1-y)}{M_i^2 - y M_{H_1}^2 - M_{H_2}^2} \qty( \log \qty[ \qty(M_i^2 - y M_{H_1}^2)(1-y)] - \log M_{H_2}^2(1-y) ) 
] \,.
\end{align}
Again, using \eqref{eq:formula2}, 
we obtain 
\begin{align}
\Im D_2 &= 
\frac{M_{H_{1}}^{2}}{16 \pi} \int_{\beta}^{1} \dd y \frac{M_{H_{2}}^{2} y(1-y)}{\left(M_i^{2}- y M_{H_{1}}^{2}  - M_{H_{2}}^{2} \right)^{2}} \\
&= \frac{1}{16 \pi M_{H_1}^4} 
\qty[
\qty( M_{H_1}^2 - M_i^2) \qty(-2 M_{H_2}^2 + M_i^2 )
- 2 M_{H_2}^2 \qty( M_{H_1}^2 + 2 M_{H_2}^2 - 2 M_i^2 ) 
\log \qty( \frac{M_{H_2}^2}{M_{H_1}^2 + M_{H_2}^2 - M_i^2} )
]\,.
\label{eq:ImD2}
\end{align}
Combining \eqref{eq:ImD1} and \eqref{eq:ImD2}, 
we obtain the imaginary part of $\mathcal{A_0^{*} \mathcal{A_1}}^{\rm (vertex)}$ as follows
\begin{align}
\operatorname{Im} \left(\mathcal{A_0^{*} \mathcal{A_1}}^{\rm (vertex)}\right)
&= M_{H_{1}}^2 \operatorname{Im}D
\\
&=  
\frac{1}{16 \pi} 
\qty[ \qty( M_{H_1}^2 - M_i^2 ) + M_{H_2}^2 \log \qty( \frac{M_{H_2}^2}{M_{H_1}^2 + M_{H_2}^2 - M_i^2} ) ]  \,.
\label{eq:Imvertex}
\end{align}
for $M_i < M_{H_1}$.

By using Eqs.~\eqref{asymmetry} and \eqref{eq:Imvertex}, we finally obtain the asymmetry from the vertex correction
\begin{equation}
\epsilon^{\rm (vertex)}_{\bar e_{f'} \ell_{f}} = -\frac{1}{8 \pi} \frac{\sum_{ij} \operatorname{Im}\left( Y_{j f'}^{2} Y_{ff'}^{1 *} \lambda_{f i}^{2} \lambda_{ji}^{1 *}\right)}{ \sum_{ff'}  \left|Y_{ff'}^{1}\right|^{2}} \frac{1}{M_{H_1}^2}
\qty[ \qty( M_{H_1}^2 - M_i^2 ) + M_{H_2}^2 \log \qty( \frac{M_{H_2}^2}{M_{H_1}^2 + M_{H_2}^2 - M_i^2} ) ]  \,.
\end{equation}
for $M_i < M_{H_1}$.

\section{Detailed calculations of CP-violating heavy Higgs decay into right-handed neutrinos}
\label{sec:appendixC}

We also show calculations for the asymmetry of right-handed neutrinos generated from the heavy-Higgs-boson decay, including the dependence on the right-handed neutrino masses, $M_i$. We adopt notations similar to the ones used in App.~\ref{sec:appB}. The calculation can be done parallelly to the previous appendix with the replacement of $\bar e_{f'} \to \bar \ell_{f}$ and $\ell_{f} \to N_i$, except that the massive fermion exists in the on-shell final state rather than in the virtual loop.

The tree diagram gives 
\begin{align}
\left| \mathcal{M}_{H_1 \to \bar \ell_{f} N_i}^{(0)} \right|^{2} 
& = - 2 \abs{\lambda_{f i}}^2 p_\mu q^\mu 
\\
&= \abs{\lambda_{f i}}^2 \qty( M_{H_1}^2 - M_i^2) 
\end{align}
where we use $q^2 = 0$ and $2 p_\mu q^\mu = -(M_{H_{1}}^2 - M_i^2)$.
The tree level decay rate is then given by 
\begin{equation}
\Gamma_{H_1 \to \bar \ell_{f} N_i}^{(0)} 
= \frac{\abs{\lambda_{f i}}^2}{16\pi} \frac{\qty( M_{H_1}^2 - M_i^2)^2}{M_{H_1}^3},  
\end{equation}

\subsection{Self-energy correction}

The diagrams for self-energy correction leads to the amplitude of 
\begin{align}
\mathcal{M}_{H_1 \to \bar \ell_{f} N_i}^{\rm (self)} 
 &= \sum_{i,j}  \bar{u}(p+q) (-i) P_{L} \lambda_{fi}^{2 *} \varepsilon_{b c} v(-q) \frac{i}{p^{2}-M_{H_{2}}^{2}} 
\nonumber\\
&\quad \times(-1) \int \frac{\dd^{4} k}{(2 \pi)^{4}}\mathrm{tr}\left[(-i) \delta_{d c} Y_{f' f''}^{2 *} P_{L} \frac{i(\slashed k+\slashed p)}{(k+p)^{2}}  (-i) \delta_{d a} Y_{f' f''}^{1} P_{R} \frac{i \slashed k}{k^{2}}\right]
\nonumber\\
 &= -2 \varepsilon_{ba} \sum_{f' f''} \lambda_{fi}^{2 *} Y_{f'f''}^{2 *} Y_{f'f''}^{1} \frac{i  }{p^{2}-M_{H_{2}}^{2}}\left( A+ p_{\mu} B^{\mu}\right)_{M_i \to 0} \bar{u}(p+q) P_{L} v(-q) \,,
\end{align}
where $A$ and $B$ are given by \eqref{eq:A} and \eqref{eq:B} with $M_i \to 0$.

The interference between the tree and self-energy is written as 
\begin{align}
\qty( \mathcal{A}_{0}^{*} \mathcal{A}^{\rm (self)}_{1} )_{H_1 \to \bar \ell_{f} N_i}
&  = - \frac{2 \qty(M_{H_{1}}^2 - M_i^2) }{M_{H_{2}}^{2} - M_{H_{1}}^{2}}  \left(A+p_{\mu} B^{\mu}\right)_{M_i \to 0}  \,. \label{treeself}
\end{align}
Substituting \eqref{eq:bB} with $M_i \to 0$, 
we obtain the asymmetry from the self-energy correction such as 
\begin{align}
\epsilon^{\rm (self)}_{\bar \ell_{f} N_i} =  -\frac{1}{8 \pi} \frac{\sum_{f' f''} \operatorname{Im}\left(\lambda_{fi}^{2 *} \lambda_{fi}^{1 } Y_{f' f''}^{2 *} Y_{f' f''}^{1 }\right)}{\sum_{f' f''} \left|\lambda_{f' f''}^{1}\right|^{2}}  \frac{M_{H_{1}}^{2}}{M_{H_{2}}^{2}-M_{H_{1}}^{2}} \,,
\end{align}
for $M_i<M_{H_{1}}$.

\subsection{Vertex correction}

The diagrams for vertex correction gives the amplitude of 
\begin{align}
\mathcal{M}_{H_1 \to \bar \ell_{f} N_i}^{\rm (vertex)}
&=  \sum_{f' f''} \bar{u}(p+q)(-i) \delta_{bc} Y_{ff'}^{2 *} \int \frac{\dd^4 k}{(2\pi)^4} \frac{i\qty[(\slashed p + \slashed q + \slashed k)   ] }{(p+q+k)^2} 
\nonumber
\\
&\qquad \times (-i) \delta_{da} Y_{f''f'}^{1} \frac{i(\slashed q + \slashed k)}{(q+k)^2} (-i) \varepsilon_{dc} P_L \lambda_{f'' i}^{2 *}  \frac{i}{k^{2}-M_{H_{2}}^{2}} v(-q)
\nonumber\\
&=  \varepsilon_{ba} \sum_{f' f''} Y_{f f'}^{2 *} Y_{f'' f'}^{1 } \lambda_{f'' i}^{2 *} (-i) 
\qty[ \tilde{D} \bar{u}(p+q) P_{L}v(-q) + M_i D'_\mu \bar{u}(p+q) \gamma^\mu P_{L}v(-q) ]
\,,
\end{align}
where we use the equation of motion $\bar{u}(p+q) (\slashed p + \slashed q) = M_i \bar{u}(p+q)$ and $\slashed q v (-q) = 0$. 
Here, $\tilde{D}$ is defined by \eqref{Ddef} with $M_i \to 0$ but with $(p+q)^2 = M_i^2$. 
Also, we define
\begin{equation}
 D'_\mu 
 \equiv  -i \int \frac{\dd^{4} k}{(2 \pi)^{4}} \frac{k_\mu}{\qty[(p+q+k)^{2} ] (q+k)^{2}\left(k^{2}-M_{H_{2}}^{2}\right)}
      \label{D'} \,, 
\end{equation}
Note that 
\begin{align}
 D'_\mu q^\mu 
 &=  -i \int \frac{\dd^{4} k}{(2 \pi)^{4}} \frac{(q+k)^2 - k_\mu (q^\mu+k^\mu)}{\qty[(p+q+k)^{2} ] (q+k)^{2}\left(k^{2}-M_{H_{2}}^{2}\right)}
 \\
 &=  \frac{1}{2} D'' 
  -  \frac{1}{2} \tilde{D} 
\end{align}
where 
\begin{equation}
 D'' 
 \equiv -i \int \frac{\dd^{4} k}{(2 \pi)^{4}} \frac{1}{\qty[(p+q+k)^{2} ] \left(k^{2}-M_{H_{2}}^{2}\right)}
\end{equation}
The finite part of $D''$ is 
\begin{align}
    D'' \to - \frac{1}{16 \pi^{2}} \int_{0}^{1} \dd x \log \qty( M_{H_2}^2 x - M_i^2 (1-x)x ) \,,
    \label{eq:self-pB2}
\end{align}
which is real for $M_{H_2} > M_i$ and does not contribute to the asymmetry.

We note that  
the interference between tree and vertex gives 
\begin{align}
&\qty( \mathcal{A}_{0}^{*} \mathcal{A}^{\rm (vertex)}_{1} )_{H_1 \to \bar \ell_{f} N_i}
\nonumber\\
& =\sum_{s,r} \sum_b \bar{v}^r(-q) \varepsilon_{ba} P_{R} i  u^s(p+q) \varepsilon_{ba} (-i) 
\qty[ \tilde{D} \bar{u}^s(p+q) P_L v^r(-q) + M_i D'_\mu \bar{u}^s(p+q) \gamma^\mu P_L v^r(-q) ] \\
& = \tilde{D} \qty(M_{H_{1}}^2 - M_i^2) 
- 2 M_i^2 D'_\mu q^\mu \,.
\\
& =  M_{H_{1}}^2 \tilde{D}
- M_i^2 D'' \,.
\label{treevertex}
\end{align}

To calculate the imaginary part of $\tilde{D}$, one has to remind of $(p+q)^2 = M_i^2$ and $2 p \cdot q = M_i^2 -M_{H_1}^2$ in the present case. It can be written as 
\begin{align}
    \tilde{D} &\equiv  -i \int \frac{\dd^{4} k}{(2 \pi)^{4}} \frac{k^2}{\qty[(p+q+k)^{2} ] (q+k)^{2}\left(k^{2}-M_{H_{2}}^{2}\right)}
 \,, 
\\
&= -i \int \frac{\dd^{4} l}{(2 \pi)^{4}} \int_0^1 \dd y \int_0^{1-y} \dd x \frac{2\left( l^2- (M^2_{H_{1}} - M_i^2) xy + x^2 M_i^2 \right)}{\left(l^2-\tilde{\Delta}^{\prime} \right)^3} 
      \\
      &\equiv  \tilde{D}_1 + \tilde{D}_2 + \tilde{D}_3 
      \label{D} \,, 
\end{align}
with 
\begin{align}
    &\tilde{\Delta}^{\prime} \equiv (y M_i^2 -yM_{H_{1}}^2-M_{H_{2}}^2)x-M^2_{H_{2}}y+M_{H_{2}}^2 \,. 
    \\
    &\tilde{D}_1 \equiv -i \int \frac{\dd^{4} l}{(2 \pi)^{4}} \int_0^1 \dd y \int_0^{1-y} \dd x \frac{2 l^2}{\left(l^2-\tilde{\Delta}^{\prime} \right)^3} 
    \\
    &\tilde{D}_2 \equiv -i \int \frac{\dd^{4} l}{(2 \pi)^{4}} \int_0^1 \dd y \int_0^{1-y} \dd x \frac{-2 (M^2_{H_{1}} - M_i^2) xy}{\left(l^2-\tilde{\Delta}^{\prime} \right)^3} 
    \\
    &\tilde{D}_3 \equiv -i \int \frac{\dd^{4} l}{(2 \pi)^{4}} \int_0^1 \dd y \int_0^{1-y} \dd x \frac{2 M_i^2 x^2}{\left(l^2-\tilde{\Delta}^{\prime} \right)^3}
    \,.
\end{align}

The imaginary part reads 
\begin{align}
\Im \tilde{D}_1 &= 
\frac{1}{8 \pi} \int_0^{1} \dd y \frac{ y M_{H_{1}}^{2} - yM_i^{2}}{y M_{H_{1}}^{2} +M_{H_{2}}^{2}-y M_i^{2}} (1-y) \\
&= \frac{1}{16 \pi (M_{H_1}^2 - M_i^2)^2} 
\Bigg[(M_{H_1}^2 - M_i^2) (M_{H_1}^2 + 2 M_{H_2}^2 - M_i^2) \nonumber \\[-.5em]
        & \qquad \qquad \qquad \qquad \qquad - 2 M_{H_2}^2 (M_{H_1}^2 + M_{H_2}^2 - M_i^2)  \log \qty(\frac{M_{H_1}^2 + M_{H_2}^2 - M_i^2}{M_{H_2}^2})
    \Bigg]\,.
\end{align}
and
\begin{align}
\Im \tilde{D}_2 &= 
\frac{M_{H_{1}}^{2} - M_i^2}{16 \pi} \int_0^{1} \dd y \frac{M_{H_{2}}^{2} y(1-y)}{\left(y M_i^{2}- y M_{H_{1}}^{2}  - M_{H_{2}}^{2} \right)^{2}} \\
&= \frac{M_{H_2}^2}{16 \pi (M_{H_1}^2 - M_i^2)^2} 
\qty[
-2 M_{H_1}^2 + 
   2 M_i^2 + (M_{H_1}^2 + 2 M_{H_2}^2 - M_i^2) 
\log \qty(\frac{M_{H_1}^2 + M_{H_2}^2 - M_i^2}{M_{H_2}^2})
]\,.
\label{eq:ImD22}
\end{align}
Also we obtain 
\begin{align} \label{last D}
\tilde{D}_3 &= 
\frac{ -M_i^{2}}{16 \pi^{2}} \int_{0}^{1} \dd x  \int_0^{1-x} \dd y 
\frac{x^2}{y (x M_i^{2}- x M_{H_1}^2 - M_{H_{2}}^{2}) + M_{H_2}^2 (1+x) }
\\
&=
\frac{ -M_i^{2}}{16 \pi^{2}} \int_{0}^{1} \dd x   
\frac{x^2}{x M_i^{2}- x M_{H_1}^2 - M_{H_{2}}^{2} } 
\log \qty[
\frac{(1-x) (x M_i^{2}- x M_{H_1}^2 - M_{H_{2}}^{2}) + M_{H_2}^2 (1+x)
}{M_{H_2}^2 (1+x)}
]
\,.
\end{align}
This does not give an imaginary part. 
Combining these, we obtain 
\begin{align}
\Im \tilde{D} &= 
 \frac{1}{16 \pi (M_{H_1}^2 - M_i^2)} 
\qty[
M_{H_1}^2 - M_i^2 - M_{H_2}^2  \log \qty(\frac{M_{H_1}^2 + M_{H_2}^2 - M_i^2}{M_{H_2}^2})
]\,.
\end{align}

We finally obtain the asymmetry from the vertex correction such as 
\begin{align}
\epsilon^{\rm (vertex)}_{\bar \ell_{f} N_i} =  - \frac{1}{8 \pi} \frac{\sum_{f' f''} \operatorname{Im}\left(
Y_{f f'}^{2 *} Y_{f'' f'}^{1 } \lambda_{f'' i}^{2 *} \lambda_{f i}^{1 }
\right)}{\sum_{f' f''} \left|\lambda_{f' f''}^{1}\right|^{2}}  \frac{M_{H_{1}}^{2}}{(M_{H_{1}}^{2}-M_i^{2})^2} 
\qty[
M_{H_1}^2 - M_i^2 - M_{H_2}^2  \log \qty(\frac{M_{H_1}^2 + M_{H_2}^2 - M_i^2}{M_{H_2}^2})
]
\,,
\end{align}
for $M_i<M_{H_{1}}$.

\bibliographystyle{utphys}
\bibliography{ref}

\end{document}